\documentclass[12pt,preprint]{emulateapj}
\begin{document}
\title{The Structure of Active Merger Remnant NGC 6240 from IRAC Observations }
\author{Stephanie J. Bush\altaffilmark{1}, Zhong Wang\altaffilmark{1}, Margarita Karovska\altaffilmark{1} and Giovanni G. Fazio\altaffilmark{1}}
\altaffiltext{1}{Harvard-Smithsonian Center for Astrophysics, 60 Garden St, Cambridge, MA 02143 USA}

\email{sbush@cfa.harvard.edu}


\slugcomment{Accepted for publication in ApJ}

\begin{abstract}

NGC 6240 is a rare object in the local universe: an active merger remnant viewed at the point of merging where two active galactic nuclei are visible. We present IRAC data of this object, providing high sensitivity maps of the stellar and PAH distribution in this complicated system. We use photometry to analyze the variation in these distributions with radius and provide an SED in the four IRAC bands: 3.6\,$\mu$m, 4.5\,$\mu$m, 5.8\,$\mu$m and 8\,$\mu$m. We fit the radial profiles of the 3.6\,$\mu$m band to $r^{\frac{1}{4}}$ and exponential profiles to evaluate the structure of the remnant. Finally, we compare the IRAC images with multi-wavelength data and examine how outflows in the X-ray, H$\alpha$ and CO correlate with 8\,$\mu$m emission. The results support the general picture of NGC 6240 as a system experiencing a major merger and transitioning from a disk galaxy to a spheroid. The sensitivity of IRAC to low surface brightness mid-infrared features provides detailed information on the extended distributions of stars and dust in this rare system.

\end{abstract}

\keywords{ galaxies: photometry,  galaxies: active,  galaxies: irregular,  infrared: galaxies,  galaxies: interactions}

\section{Introduction}

 Due to its proximity, NGC 6240 ($z=.0243$ \citet{Solomon-et-al-1997}, $D=98$ Mpc for $H_{0}=75$ km s$^{-1}$ Mpc$^{-1}$) is one of the brightest objects in the IRAS All-Sky Survey and was one of the early identified LIRGs with a total infrared luminosity of $7.1\times10^{11} L_{\odot}$ \citep{Wright-et-al-1984}.   Optical
studies of NGC 6240 reveal a morphologically irregular galaxy
with clear tidal extensions, two nuclei
and large dust lanes, strongly suggesting a merging system \citep{Zwicky-et-al-1961, Fried-Schulz-1983, Keel-1990, Gerssen-et-al-2004}. Near-IR data \citep{Thronson-et-al-1990, Doyon-et-al-1994, Scoville-et-al-2000, Max-et-al-2005} and radio continuum \citep{Carral-et-al-1990, Eales-et-al-1990, Colbert-et-al-1994, Beswick-et-al-2001} images confirm the double nuclei and show that they have considerable substructure. The observed distance between the the nuclei  depends on the wavelength at which they are observed, indicating large amounts of dust extinction in the center of the galaxy \citep{Schulz-et-al-1993, Gerssen-et-al-2004}. H$\alpha$ data shows strong
evidence for nuclear winds, outflows and bubbles
\citep{Gerssen-et-al-2004}.  X-ray studies show a
very active galaxy with two nuclei
displaying AGN characteristics, indicating two
black holes \citep{Iwasawa-Comastri-1998, Vignati-et-al-1999, Komossa-et-al-2003}. Sub-millimeter
array data reveal a molecular gas disk of mass $\approx 10^{9}$ M$_{\odot}$ in between
the two nuclei with the possibility of outflows \citep{Wang-Scoville-Sanders-1991, Tacconi-et-al-1999, Iono-et-al-2007}.  Merger remnants with double nuclei are rare in the nearby universe, suggesting that we are catching this object in a very short lived phase of its evolution. In addition, its infrared luminosity places it on border between the LIRG and a ULIRG classes. Though classified by their luminosity, the LIRG and ULIRG classes show differing characteristics besides their luminosity, including a higher occurrence of AGN and disturbed morphologies in ULIRGs \citep{Veilleux-et-al-2002, Sanders-et-al-2004}. Given that NGC 6240 already shows many of these features, we may be observing a transition object between the LIRG and ULIRG classes. 

Objects such as NGC 6240 are believed to be undergoing a merger between two comparable mass galaxies which is driving the
remnant through evolution from a
Seyfert type active galaxy to an AGN and finally a quasar \citep{Norman-Scoville-1988, Sanders-et-al-1988, Hopkins-et-al-2006}. Models
demonstrate a cycle in which a merger drives gas
inflows that create starbursts and feed
central super-massive black holes, triggering an
active stage for the black hole. As the AGN
``turns on'' and contributes significant
radiative feedback, it clears out the gas around
the nuclei, and the AGN becomes visible as a
quasar. However, this also depletes its fuel,
leaving the AGN to die back to a quiescent phase.
The cycle may start over again when the galaxy goes
through another merger. \citet{Hopkins-et-al-2008-redgals, Hopkins-et-al-2008-quasars} show this cycle is consistent with many observed properties of quasars and elliptical galaxies. This evolutionary sequence is short ($\sim$ few Gyr) compared to the lifetime of a spiral or elliptical galaxy, so there are relatively few of them to study, and even fewer at the particular stage of merging where two AGN are observable. This makes them particularly interesting objects to study in detail. 

We add
Spitzer/IRAC \citep{Fazio-et-al-2004} data of NGC
6240 to the existing observations of NGC 6240. Most infrared studies of NGC 6240 have focused on
the nuclei \citep{Klaas-et-al-2001, Lutz-et-al-2003, Armus-et-al-2006, Egami-et-al-2006}. We draw on IRAC's high
sensitivity and multiple mid-infrared bands to investigate how stars and dust
are distributed throughout the remnant. The IRAC band-passes single out emission from stars (3.6\,$\mu$m) and polycyclic aromatic hydrocarbon emission (PAH, a component of dust, 8\,$\mu$m), allowing us to dissect the structure of NGC 6240 by examining the distribution of stars and ISM separately \citep{Pahre-et-al-2004}. All four IRAC bands are imaged on the same scale so that these components can be easily compared. We derive light profiles in each of the IRAC bands and calculate the SED. We also compare IRAC data to archive optical, X-ray and radio data. In \S\ref{sec:obs} we describe our observations and in \S\ref{sec:results} we describe the IRAC images and how they compare to data from other wavelengths. In \S\ref{sec:discuss} we present our photometry of the IRAC data and in \S\ref{sec:conc} we make some concluding remarks.

\section{Observations} \label{sec:obs}

Observations were carried out with IRAC \citep{Fazio-et-al-2004} as part of the Spitzer GTO program on 13 August 2004 in the 3.6, 4.5, 5.8 and 8.0\,$\mu$m bands. Because of NGC 6240's high infrared luminosity, the data was taken in high dynamic mode. Sixty dithered images with an exposure time of 0.6 s were taken in each filter in addition to  120 dithered images with an exposure time of 10.4 s. The short exposure images allow checks against saturation and bright source effects. The short and long exposures were mosaic-ed separately giving total exposure times of 10.8 mins and 36 s in the long and short exposure mosaics, respectively. The IRAC post pipeline Basic Calibrated Data (BCDs, pipeline version 14.0.0) were corrected for artifacts with an artifact mitigation code \citep{Carey-2007} and were mosaiced with IRACproc (v14.1, \citet{Schuster-et-al-2006}). The pixel size of the BCDs is 1.22$\arcsec$. The pixel size of the mosaiced images is 0.86$\arcsec$, this is the minimum
re-sampling needed to ensure that the IRAC PRF is not under-sampled even at 3.6
and 4.5 $\mu$ms. The FWHM of the IRAC point response functions are all $\sim 2.0 \arcsec$, which corresponds to 950 pc at the distance of NGC 6240. The double AGN of NGC 6240 are $\sim$ 1 kpc apart \citep[e.g.][]{Gerssen-et-al-2004}, so we cannot resolve them and will concentrate on the extended emission of the remnant. 

We compare the IRAC data to archival data from many sources including Chandra Soft X-ray, Hubble Space Telescope Advanced Camera for Surveys (ACS), Very Large Array (VLA) radio continuum and sub-millimeter array  (SMA) CO 3-2.  The Chandra observations of NGC 6240 were carried out over
37 ks on 2001 July 29 (OBSID 1590) using the ACIS-S instrument.
The pixel size of the ACIS detector is 0.49".
We analyzed the data using {\it CIAO}\footnote{CIAO is the Chandra Interactive Analysis of Observation's
data analysis system package (http://cxc.harvard.edu/ciao)} data
reduction and analysis routines. 
To detect the low contrast diffuse emission surrounding the bright
 nuclear region we applied the CIAO adaptive
smoothing tool {\it csmooth} \citep{Ebeling-et-al-2006}.
{\it Csmooth} preserves the multi-scale
spatial signatures
and the associated counts \citep[e.g.][]{Karovska-et-al-2002, Karovska-et-al-2007}.
The smoothing is achieved by convolution with a
Gaussian kernel which increases from a small initial
smoothing scale
until the total number of counts under the kernel exceeds a
value that is determined from a preset significance and the
expected number of background counts in the kernel area.
The maximum smoothing scale in the image is 5".

Hubble Space Telescope ACS data in the B broadband filter and Wide Field Planetary Camera (WFPC2) H$\alpha$ \citep{Gerssen-et-al-2004} data were obtained from the NASA Multi-mission Archive at STScI (MAST)\footnote{http://archive.stsci.edu}. VLA radio continuum data centered at 21-cm was obtained from the NASA extragalactic database \citep[NED\footnote{http://nedwww.ipac.caltech.edu/};][]{Condon-et-al-1996} and sub-millimeter array data of CO 3-2 was obtained with the Submillimeter Array \citep[SMA,][]{Iono-et-al-2007}.

\section{Results} \label{sec:results}

\subsection{IRAC Imaging} \label{sec:morph}

\begin{figure*}
\plotone{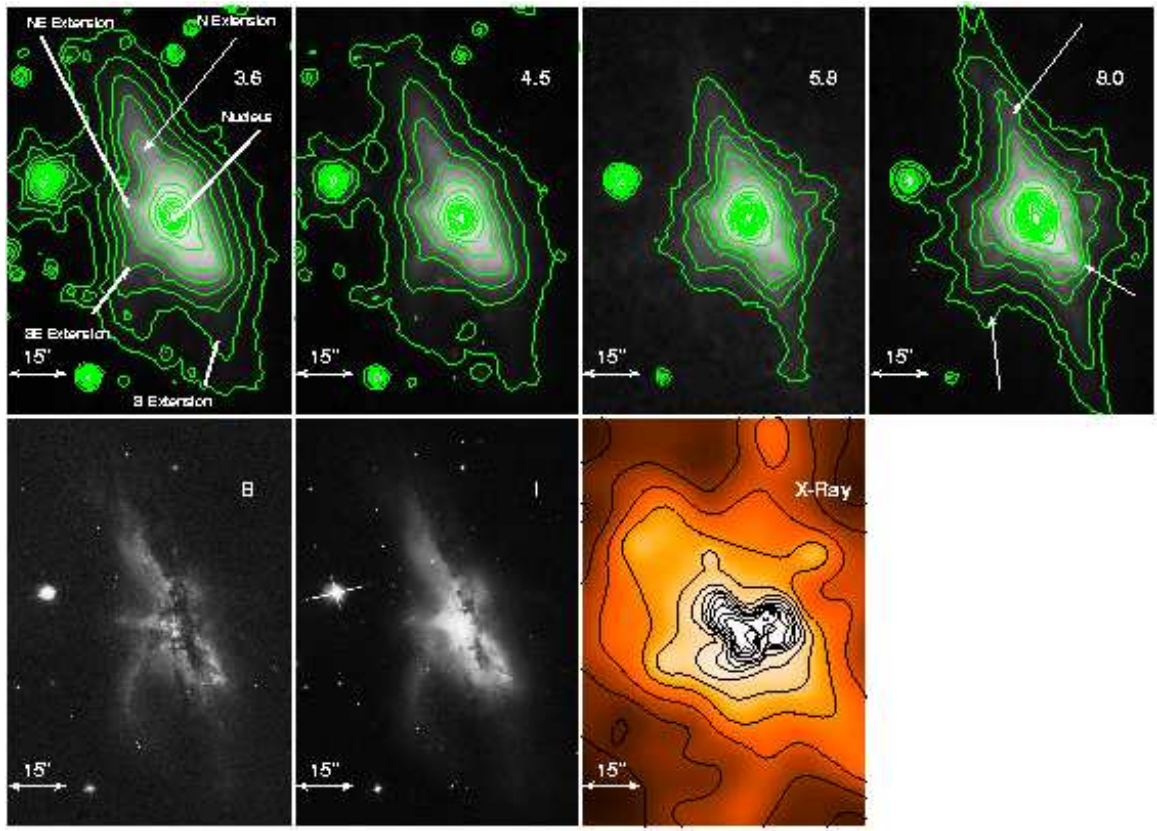}
\caption[]{Multi-wavelength data of NGC 6240. On top are the four IRAC broadband images with contours. North is up and east is to the left. In all images the contours are logarithmically spaced with the lowest contour at 4$\sigma$ above the background. The minimum contours are .16, .33, 5.06, 9.68 MJy/str and maximum contours are  68.79, 88.59, 249.2, 400.1 MJy/str for the four bands respectively.  In the 5.8\,$\mu$m images additional contours are added between the first two contours at 5.5, 6, and 6.5 MJy/str. In the 8.0\,$\mu$m image additional contours are added between the first two contours at 10, 11 and 12 MJy/str. A 15\arcsec scale is shown in the image. At the distance of NGC 6240, this corresponds to 7.1 kpc. Note that the 5.8 and 8.0\,$\mu$m bands show prominent PRF effects, see the PRF in Figure~\ref{fig:8prf}. On the bottom are ACS B, I and Chandra soft X-ray images of NGC 6240 at the same scale. Logarithmic contours are shown on the X-ray image.}
\label{fig:IRAC4chan}
\end{figure*}

\begin{figure}
\plotone{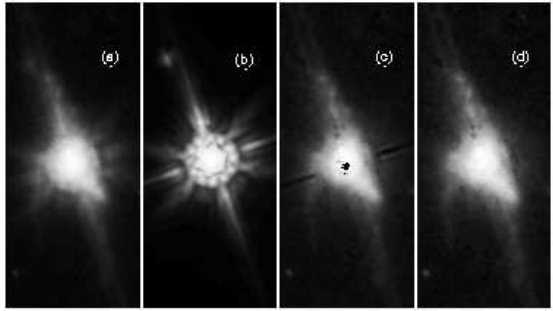}
\caption[]{This figure illustrates how we processed the 8\,$\mu$m image to remove the central PRF. (a) The original 8\,$\mu$m image. (b) The point response function of the  8.0\,$\mu$m band showing the PRF spikes that afflict the 8\,$\mu$m image. This PRF is saturated to show the outer PRF features, so these effects will be significantly fainter in the image of NGC 6240.  (c) The 8\,$\mu$m image with the PRF subtracted out. Because this is not a point source, the subtraction cannot be done perfectly, and some artifacts are left at the center of the image. The lines at the side are a well known IRAC artifact. (d) Finally, these artifacts are cleaned to make the image more visually pleasing. No photometry is done on this image, but it is used for creating all the following images.}
\label{fig:8prf}
\end{figure}

In Figure~\ref{fig:IRAC4chan} we show the IRAC images in each bandpass along with the ACS B-band, ACS I-band and Chandra soft X-ray images. Contours are overlaid on the IRAC images to highlight structure. The lowest contour is set at $4\sigma$ above the background, which is .16, .33, 5.06, 9.68 MJy/str in the 3.6, 4.5, 5.8 and 8.0\,$\mu$m bands respectively. Four features extending off the remnant are labeled in the 3.6\,$\mu$m panel of Figure~\ref{fig:IRAC4chan}. They point approximately south, north, southeast (then curves around to the south) and northeast (smaller) and are labeled as such.  The 8.0\,$\mu$m image is dominated by ``spikes'' of emission to the east and west of the galaxy. These are due to the bright central source illuminating the IRAC 8.0\,$\mu$m point response function and are not true features. They are clearly seen in the PRF of the 8.0\,$\mu$m band shown in Figure~\ref{fig:8prf}, panel (b). There is also some contribution from the PRF to the 8\,$\mu$m image along the north-south axis of the remnant.  Three low intensity features that are not due to the PRF are indicated with arrows on the 8\,$\mu$m panel in Figure~\ref{fig:IRAC4chan}: a dust compliment to the southeastern feature, a bright source on the southwest side of the remnant and the dust lane to the north. Finally, note that the bright source to the northeast of the remnant is a foreground star. 

In an attempt to remove the PRF artifacts and clarify the dust features in the 8\,$\mu$m image, we subtract off the PRF of the central emission, treating it as a point source (Figure~\ref{fig:8prf}). The magnitude of the subtracted object is $\sim$ 6 (Vega). A few artifacts are left in the center after the subtraction, and these pixels are set to an average value of the pixels around them. In addition, the reflection artifacts shown on either side of the remnant are cleaned by averaging over columns of bright and dark pixels. This means that for these six pixels on either side of the remnant, features are smoothed in the NE-SW direction. The final image is shown in Figure~\ref{fig:8prf}d. Features far from the central source of the remnant should be largely unaffected by remaining artifacts, but any remaining axisymmetric features should not be trusted. This image is never used for photometry, but it allows us to exploit IRAC's strengths to examine the morphology of the extended emission in NGC 6240. Figure~\ref{fig:8prf} shows the original 8\,$\mu$m image, the 8\,$\mu$m PRF and the final image. Dust features along the northern, southern and southeastern extensions are now clearly visible and show noticeable substructure. The northern extension in particular appears very clumpy. The southeastern extension is pronounced in 8\,$\mu$m and seems to be a long, thin tidal tail. The southern extension is quite faint compared to the other features in the 8\,$\mu$m and curves such that the tip is pointing to the southeast. Along the southeastern side of the remnant, between the southern and southeastern extensions, dust filaments are present, suggestive of entrained dust being blown out of the remnant. 

\begin{figure*}
\plotone{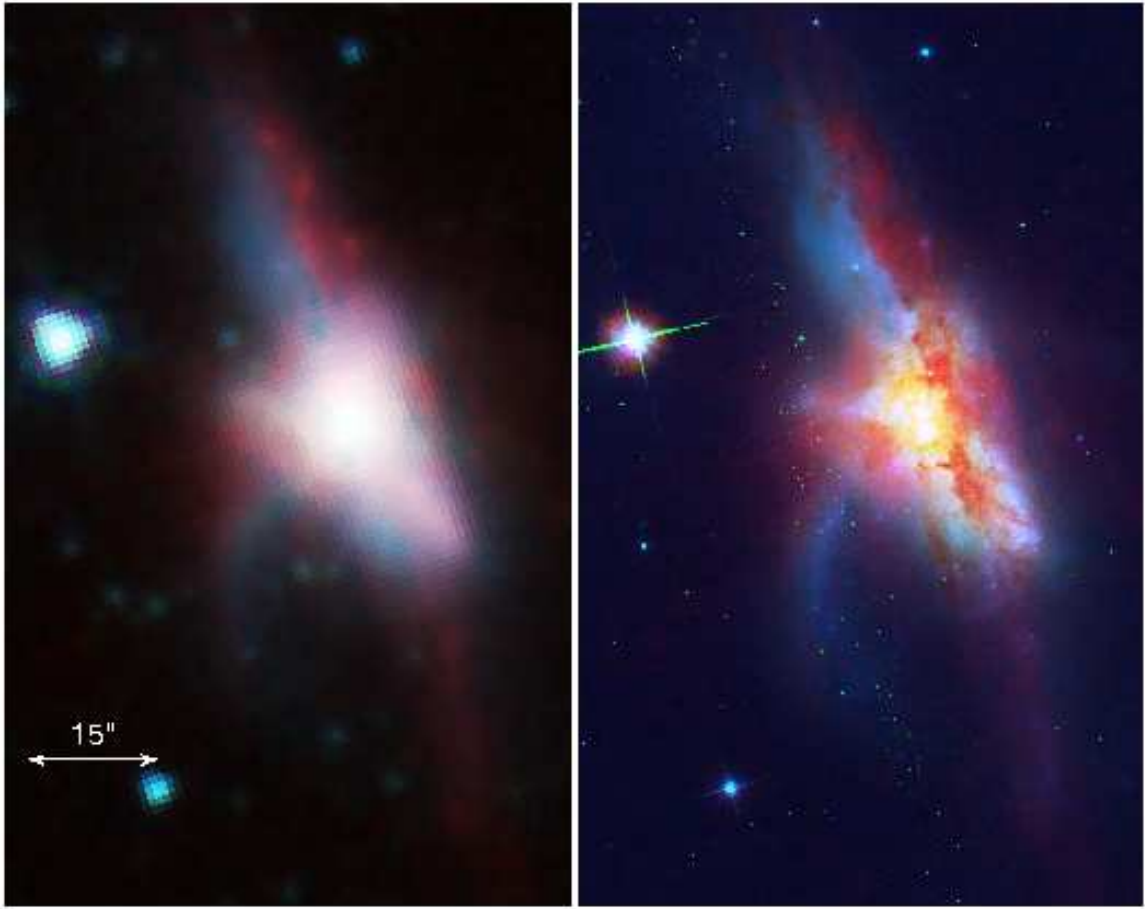}
\caption[]{Optical and IRAC three-color images. Left: IRAC 3.6\,$\mu$m in blue, IRAC 4.5\,$\mu$m in green and PRF subtracted 8\,$\mu$m in red. Right: ACS B-band in blue, ACS I-Band in green and PRF subtracted 8\,$\mu$m in red. Notice how 8\,$\mu$m emission follows the optical dust obscuration very closely.}
\label{fig:IRAC3color}
\end{figure*}

 To examine how components of the remnant are distributed, the IRAC 3.6, 4.5 and 8\,$\mu$m data are combined to create a three-color image of NGC 6240 in Figure~\ref{fig:IRAC3color} (left). The 3.6 and 4.5\,$\mu$m bands, with less than one tenth the dust obscuration of the optical, trace old stars, despite significant dust obscuration. The 7.7\,$\mu$m and 8.6\,$\mu$m  PAH features are located in the  5.8 and 8.0\,$\mu$m bands, therefore these bands trace emission from complex dust molecules. The standard IRAC coloring is used in the three-color image (3.6\,$\mu$m - blue, 4.5\,$\mu$m - green and 8.0\,$\mu$m - red) so that stellar populations appear blue, PAH emission appears red and combinations of the two appear white. The most striking aspects of the IRAC 3-color image are the differences across the northern and southeastern features - highlighting segregation of the stellar and dust distribution.   The northern extension shows clear separation of the stars and dust - stars dominating on the eastern side and the dust dominating on the western side. The southeastern extension present in the optical is clear in the 3.6\,$\mu$m, and the 8\,$\mu$m reveals a dust counterpart lining the eastern side. The long, thin, arched appearance of this extension is characteristic of a tidal tails. A gas/star offsets in tidal tails are commonly observed in merger remnants \citep[e.g.][]{Hibbard-Yun-1999, Hibbard-et-al-2005} and can be explained by the collisional nature of the gas \citep{Mihos-2001}. The right panel of Figure~\ref{fig:IRAC3color} is a three-color combination of the ACS B (blue), I (green) and PRF subtracted 8\,$\mu$m (red) images.  Note that the resolution of the ACS data is an order of magnitude better than the IRAC data. The yellow colors in the right panel correspond to areas dominated by I and 8\,$\mu$m, while the red colors indicate only 8\,$\mu$m emission is present. Young, unobscured stellar populations appear blue. Again, areas with both dust and stellar emission appear white. Since the B and I band are subject to more dust obscuration than the 3.6 and 4.5 $\mu$m bands, only reasonably unobscured stellar populations appear white in the right image, while stellar populations that suffer more dust obscuration still appear white in the left image. 

The 8\,$\mu$m emission and optical dust lane correlate amazingly well along the northern extension, down to individual clumps in the 8\,$\mu$m emission. However, the 8\,$\mu$m also reveals dust we would not have seen in the optical, such as the dust/gas companion to the SE tail. 

More subtle features are revealed in the three-color images (Figure~\ref{fig:IRAC3color}) as well. A faint southern dust extension, revealed by the 8\,$\mu$m, is spatially coincident with the southern stellar extension in the optical and 3.6\,$\mu$m, rather than being separated as in the northern and southeastern extensions. This extension is very irregular, filamentary and truncated. Filamentary 8\,$\mu$m emission is also present along the southeast side of the remnant, between the southeastern and southern extensions. This is most clear in Figure~\ref{fig:8prf}, panel d. Nothing like this emission is seen in the shorter IRAC wavelengths, indicating this filamentary dust is unaccompanied by stars. Similar filaments are seen in M 82 \citep{Engelbracht-et-al-2006}. The northeastern extension also seems to be dominated by dust, especially along its southern edge.  

 Overall, the stellar distribution is much smoother than the dust distribution.  This may be a combination of two effects: differing distributions of gas and stars in the pre-merger galaxies and the fact that physical processes in mergers affect stars and gas differently. Notice, in Figure~\ref{fig:IRAC4chan}, that the 3.6\,$\mu$m contours are evenly spaced and straight on the east and west side of the body of the remnant, while the dust obscuration of optical light creates the irregular ``bow-tie'' appearance of this remnant.  The 8\,$\mu$m emission shows particular elongation along the southwestern part of the bow-tie, rather than following the smooth distribution of the stars. Each of the extensions is more clumpy in dust emission than in stellar emission.  Along the northern extension and the body of the remnant, where the morphology of the remnant looks like an edge-on disk, the separation of stars and dust could just be a geometric effect (edge-on disks show a prominent dust lane running along their major axis due to dust obscuration which appears off center if the disk is tilted towards or away from the observer). We use photometry to test this in Section~\ref{sec:discuss}. However, in many parts of the remnant, such as the filamentary emission in the southeastern part of the remnant, the different distributions are likely to reflect the fundamental fact that mergers affect gas differently than stars, due to its collisional nature. The clumpy and uneven dust distribution is a clear prediction of starburst and AGN feedback \citep[e.g][]{Springel-et-al-2005}.

\subsection{Comparison to Multi-Wavelength Data} 

\begin{figure}
\plotone {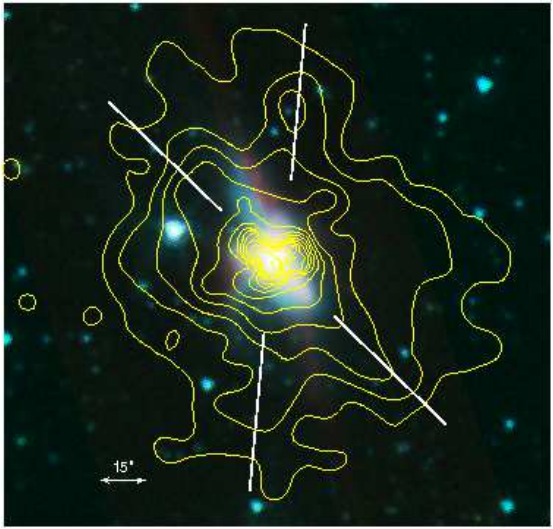}
\caption[]{X-ray contours (yellow) overlaid on IRAC three-color image. The first contour is at the 3$\sigma$ level, and they continue logarithmically. Lines are drawn to draw attention to the ``X'' pattern in the X-ray emission.}
\label{fig:xray-irac}
\end{figure}

We compile optical, soft X-ray, H$\alpha$, CO and radio data of NGC 6240 for comparison with the IRAC data. Figure~\ref{fig:xray-irac} shows the contours from the smoothed soft X-ray image (0.5-1.5 keV) overlaid on the IRAC data where the extended diffuse emission was the most prominent ($\sim$ 1000 total counts).  Soft X-rays trace hot gas ($> 10^{6}$ K). The X-ray data shows an X-like structure in the diffuse emission 
extending to $\sim$1 arcminute from
the central region with an
axis of symmetry along the north-south features of NGC 6240. The northeast-southwest line of the X is especially prominent. This pattern is probably associated with stellar outflows generated by the starburst in NGC 6240; soft X-ray emission from similar stellar outflows in starbursts is well documented in cases like M 82 \citep{Griffiths-et-al-2000}. Some of the emission may be scattered X-ray light, as in the case of Centaurus-A \citep{Karovska-et-al-2002}. In the inner regions the X-ray shows a ``cloverleaf'' pattern with three clear bubbles, most likely associated with outflows generated by the AGN \citep{Komossa-et-al-2003}. The H$\alpha$ data follows the inner X-ray contours closely and appears filamentary \citep{Gerssen-et-al-2004}, another good indication of outflows in this remnant. 

\begin{figure}
\plotone{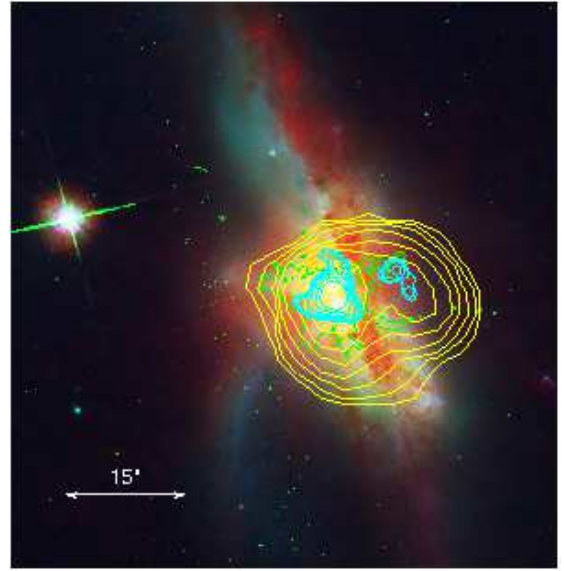}
\caption[]{The inner regions of NGC 6240 with contours showing different elements of the interstellar medium overlaid on the optical three-color image from Figure~\ref{fig:IRAC3color}. The 3$\sigma$ H$\alpha$ contour is shown in green, while CO contours are shown in cyan and radio continuum contours are shown in yellow. We only plot CO peaks considered significant by \citet{Iono-et-al-2007}.}
\label{fig:outflows}
\end{figure}

CO (tracing molecular gas) and radio continuum emission (roughly tracing synchrotron emission) contours are shown in Figure~\ref{fig:outflows}. The 3$\sigma$ contour for the H$\alpha$ emission is shown here in green and CO data is shown in cyan. We only show CO emission peaks that \citet{Iono-et-al-2007} consider significant. Radio emission is concentrated in the center, but also shows a peak arcing to the west. All three forms of gas emission show a peak over the central source, but the CO and radio data also show a peak inside a bubble outlined by the west two filaments of H$\alpha$ emission. This bubble is also coincident with the west ``leaf'' of the ``cloverleaf'' of X-ray emission shown in Figure~\ref{fig:xray-irac} \citep{Komossa-et-al-2003}. This has been interpreted as a large outflow (hereafter referred to as the western outflow), with evidence of the colder molecular gas being entrained \citep{Komossa-et-al-2003, Gerssen-et-al-2004,Iono-et-al-2007}.

While the stellar component traced by IRAC is collisonless, and will not be affected by these outflows, the dust may be affected. The 8\,$\mu$m emission clearly peaks over the central source, but does not seem to show a corresponding peak over the western outflow seen in the H$\alpha$, CO and radio. PAH emission in the western outflow region seems smooth, with no evidence for a peak or blowout. However, the PRF of the central source and effects from the PRF subtraction could affect our ability to see a peak or minimum in this region. The filamentary emission along the southeastern side and the patchy, almost filamentary emission coincident with stars in the southern extension are tantalizing indications of dust entrainment in outflows. It is possible that the temperature and pressures in the western outflow are so high, as shown by the X-ray emission there, that complex PAH molecules have been destroyed, while they survived in other, less hot outflows in NGC 6240. However, the CO peak in the western outflow shows some cold gas survives here, which we would expect to contain complex molecules such as PAHs. Other bubbles in the X-ray and H$\alpha$ seem to correspond well with the PAH emission. The filamentary emission in the southeastern part of the remnant seems to follow well from the southern H$\alpha$ filaments and southern ``leaf'' of the X-ray ``cloverleaf''. The strong PAH emission in the northeastern extension is coincident with the northeast ``leaf'' of the X-ray ``cloverleaf'' and the northeast filaments of the H$\alpha$ emission. 

\section{Analysis and Discussion} \label{sec:discuss}

\subsection{Elliptical Photometry} \label{sec:phot} 

To quantify the distribution of stars and dust in NGC 6240, IRAC colors were determined by fitting elliptical isophotes to the remnant and examining the enclosed light to investigate changes in color with radius. Using the task ``ellipse'' in IRAF, elliptical isophotes were fit to the 3.6\,$\mu$m data and used as apertures for analysis of the 4.5, 5.8 and 8\,$\mu$m bands. Removal of the 8\,$\mu$m PRF artifacts quantitatively is difficult due to the fact that the bright central source of NGC 6240 is not a point source, so photometry was performed on the original image. To compare at identical angular resolution, each of the images was convolved with a kernel \citep{Gordon-2007} which matched their PRF to the 8.0\,$\mu$m PRF before the photometry was applied.  This corrects the image for the dominant central effects (radius less than $\approx$ 4 kpc). The outer PRF ``spikes'' contribute an error of $\sim$2\% to the enclosed flux in the outer aperatures.

The semi major axis of the innermost ellipse was 0.94 kpc ($\sim 2$\arcsec) and it was increased by 0.94 kpc for each larger ellipse, which corresponds to the FWHM of the point response function of the 8\,$\mu$m band. For each ellipse, the ellipticity, position angle and central coordinates of the ellipses were allowed to vary at the predetermined semi-major axis. The ellipticity of the ellipses increases slowly from an initial value of .25 to .5 at 8 kpc, then remains approximately constant. The position angle of the ellipses varies between 80 and 90 degrees, so the semi-major axis is approximately along the remnant's north-south extensions. The maximum radius was the largest radius ellipse fit before the neighboring foreground star was enclosed by the ellipse: 20.75 kpc (44\arcsec).  Despite NGC 6240's irregular morphology in the optical, fairly concentric ellipses were fit, reflecting the smoother emission in the 3.6\,$\mu$m band than in the optical. These apertures were then used to evaluate the brightness as a function of radius in all four bands.  A constant background was subtracted from all channels. The background was calculated by taking a 115 kpc by 60 kpc box around the remnant and running 3$\sigma$ clipping iterations until the mean background value converged. The dominant error is in the ellipticity of the fitted ellipses. Uncertainties in the fitted ellipses were determined by the ``ellipse'' task from the uncertainties in the harmonic fits and final error bars were determined by changing the ellipticity of the ellipse by this error and measuring the change in the intensity enclosed by the ellipse. Statistical errors due to pixel noise are negligible. Errors due to the calibration of the original BCDs are discussed in the IRAC data handbook. As a guide, IRAC photometry is only accurate to a few percent due to calibration errors, though this will only affect absolute fluxes. Finally, to check for saturation in the 8\,$\mu$m band, we performed the same analysis on the short exposure mosaic and found virtually identical results in the inner regions. 

\begin{figure}
\plotone {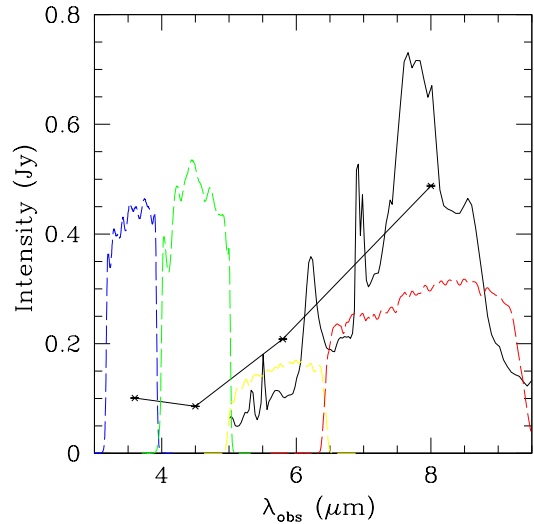}
\caption[]{ The integrated spectral energy distribution of NGC 6240 in the four IRAC bands. For comparison, IRS data is plotted for wavelengths longer than 5.0\,$\mu$m \citep{Armus-et-al-2006}. The IRS data is not scaled. The IRAC bandpasses are also plotted as the colored lines (red - 3.6, green - 4.5, yellow - 5.8 red - 8.0 $\mu$m) where the scale is correct but the units are number of electrons per incoming photon.}
\label{fig:iracsed}
\end{figure}

Using the largest elliptical aperture described we measured the total flux of NGC 6240 to be .10, .09, .21 and .49 Jy in the 3.6, 4.5, 5.8 and 8\,$\mu$m bands respectively. This SED is shown in Figure~\ref{fig:iracsed} with the Spitzer/Infrared Spectrograph (IRS) data for NGC 6240 obtained by \citet{Armus-et-al-2006}. Though this IRS data only covers wavelengths longer than 5.0\,$\mu$m, it agrees with the IRAC SED to within .1 Jy at these wavelengths. The IRAC SED also agrees well with the integrated ISO spectra presented by \citet{Lutz-et-al-2003}. Note that bands 3 and 4 show strong PAH features, indicating the overall SED is weighted strongly towards the central dust-dominated regions of the remnant. 

\begin{figure}
\plotone {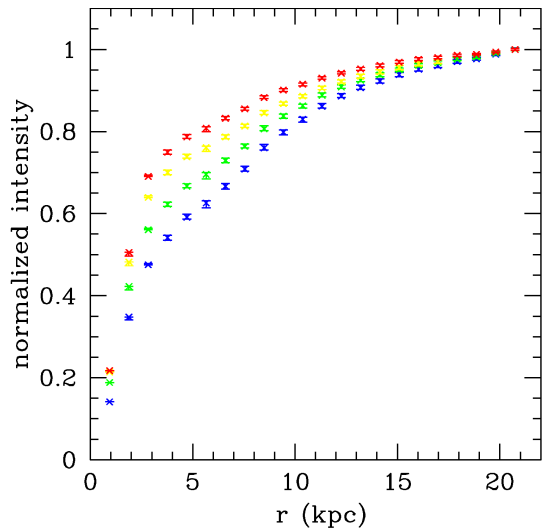}
\caption[]{Enclosed flux at a given radius derived from fitted isophotal ellipses for NGC 6240: blue - 3.6\,$\mu$m, green - 4.5\,$\mu$m, yellow - 5.8\,$\mu$m and red - 8.0\,$\mu$m. Curves are normalized to the total flux of the remnant in each band. Note that at longer wavelengths, the emission is more concentrated. }
\label{fig:ellipsephot}
\end{figure}

The flux as a function of radius for each band is shown in Figure~\ref{fig:ellipsephot}. The total flux in each band slowly levels off at large radii, indicating that the largest aperture encloses most of the flux from the galaxy, and each curve is normalized to the total flux in that band.  The longer wavelength bands gain flux much more rapidly in the central few kpc than the shorter wavelength bands, which indicates that the flux from dust emission in the galaxy is more centrally concentrated than that from stellar emission. A good way of examining the changing colors in the remnant is to calculate the ratios of the 5.8 to 3.6\,$\mu$m colors and the 8.0 to 4.5\,$\mu$m colors, effectively calculating at how red or blue the remnant is in the mid-IR colors as a function of aperture radius.  The colors calculated from the innermost aperture of our photometry, enclosing both nuclei of NGC 6240, are the reddest in 5.8/3.6\,$\mu$m of any of the apertures  and the 5.8/3.6\,$\mu$m emission always decreases with increasing aperture radius.  In 8.0/5.8\,$\mu$m emission the colors increase to a ``knee'' at 2.8 kpc, where the aperture encloses the majority of the 8\,$\mu$m emission, indicating the large concentration of dust emission in the center of the remnant. Then the colors decrease in 8.0/4.5\,$\mu$m with larger aperture.

\citet{Sajina-et-al-2005} use SED modeling to determine the color ratios of galaxies dominated by continuum, PAH and stellar sources. They find that objects dominated by different sources of emission lie in well differentiated parts of the plot, supporting observational results \citep{Lacy-et-al-2004, Hatziminaoglou-et-al-2005}, but that the populations overlap at their edges. They caution that determining the characteristics of a galaxy using this diagram is not necessarily appropriate for samples smaller than 10, due to the risk of the galaxy being an outlier, so drawing strong conclusions from NGC 6240's location in this plot would be unwise. However, in their modeling, our IRAC colors for NGC 6240 place it on the border between the PAH dominated and continuum dominated regimes, leaving the possibility that continuum emission from the AGN in NGC 6240 is seen in the IRAC colors in addition to PAH line emission.

\subsection{3.6 micron radial profile}
 
\begin{figure}
\plotone{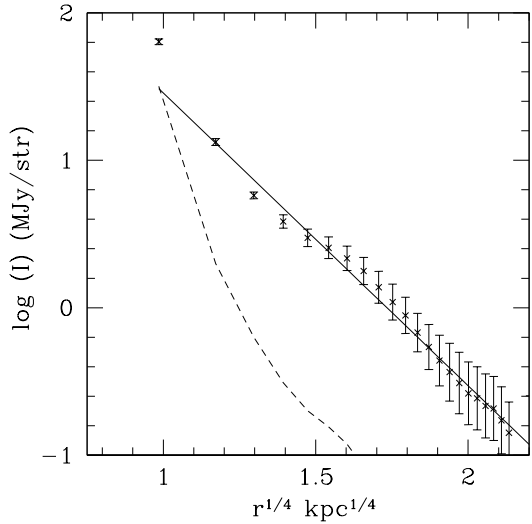}
\caption[]{The 3.6\,$\mu$m surface brightness profile derived from the elliptical isophotes plotted against $r^{1/4}$. This fit gives an effective radius of $\sim$ 8 kpc. The dashed line is the effect of the PRF.}
\label{fig:rquart}
\end{figure}

The optical morphology of NGC 6240 is highly irregular. This is the product of two effects: uneven dust obscuration and actual morphological irregularity. Our 8\,$\mu$m observations confirm the clumpy nature of the dust (see Figure~\ref{fig:IRAC3color}). With a reduction of the extinction by a factor of $\gtrsim$10 over the optical, IRAC 3.6 and 4.5\,$\mu$m data is able to penetrate the dust obscuration. To this end, we use the elliptical isophotes discussed in \S~\ref{sec:phot} to create a radial profile for NGC 6240 in 3.6\,$\mu$m to determine the true stellar distribution. We first fit an $r^{1/4}$ law, shown in Figure~\ref{fig:rquart}, to determine how well this remnant approximates an elliptical galaxy. A similar calculation was done by \citet{Doyon-et-al-1994} in the K-band. Our results and theirs are very similar, but their K-band resolution allows them to extend the profile inside of 1 kpc. IRAC's resolution does not allow this but its sensitivity allows us to extend the profile to 25 kpc, over twice as far as \citet{Doyon-et-al-1994}. Outside of 5 kpc, the fit appears to agree well with a $r^{1/4}$ law. The profile shows some ``excess light,''  a deviation above an $r^{1/4}$ law, in the central $\approx$1.5 kpc of NGC 6240. In a sample of 51 merging galaxies, \citet{Rothberg-Joseph-2004} found that 16 showed similar indications of excess light. They propose three possible explanations. First, that this could be from a population of stars formed in a starburst triggered by dissipative collapse during the merger as is shown in the models of \citet{Mihos-Hernquist-1994}. Secondly, that this light is due to a central AGN. Finally, this light could be due to the nuclear structure of the progenitor galaxies as shown by the models of \citet{Fulton-Barnes-2001}. In the case of NGC 6240, we know that two central AGN exist, making an AGN at least part of the explanation. By comparing our profile to \citet{Doyon-et-al-1994} it is clear that our profile starts outside the peak of either nuclei, so we cannot attribute this light to either nuclei. It is still unclear whether this light is from stars that were originally in one of the galaxies or part of a young starburst population. However using numerical simulations, \citet{Hopkins-et-al-2008-extralight}, show that if extra light components are robustly separated out from the light profile of the galaxy, merger remnants ubiquitously contain a component created primarily by the merger-driven starburst, that is reflected in this extra light. Given this result, it is most likely that the ``extra-light'' seen in NGC 6240 is primarily from a starburst component resulting from the merger. This extra light component has some dependence on the progenitor galaxies initial gas fraction (though, counter-intuitively, the extra light decreases with increasing progenitor gas mass because at higher initial gas masses more gas is consumed in a starburst at first passage), but the relationship has high scatter for individual objects \citep{Hopkins-et-al-2008-extralight}. In samples examined by \citet{Hopkins-et-al-2008-extralight} this component's mass spans 3-20\%. 

\begin{figure*}
\plottwo{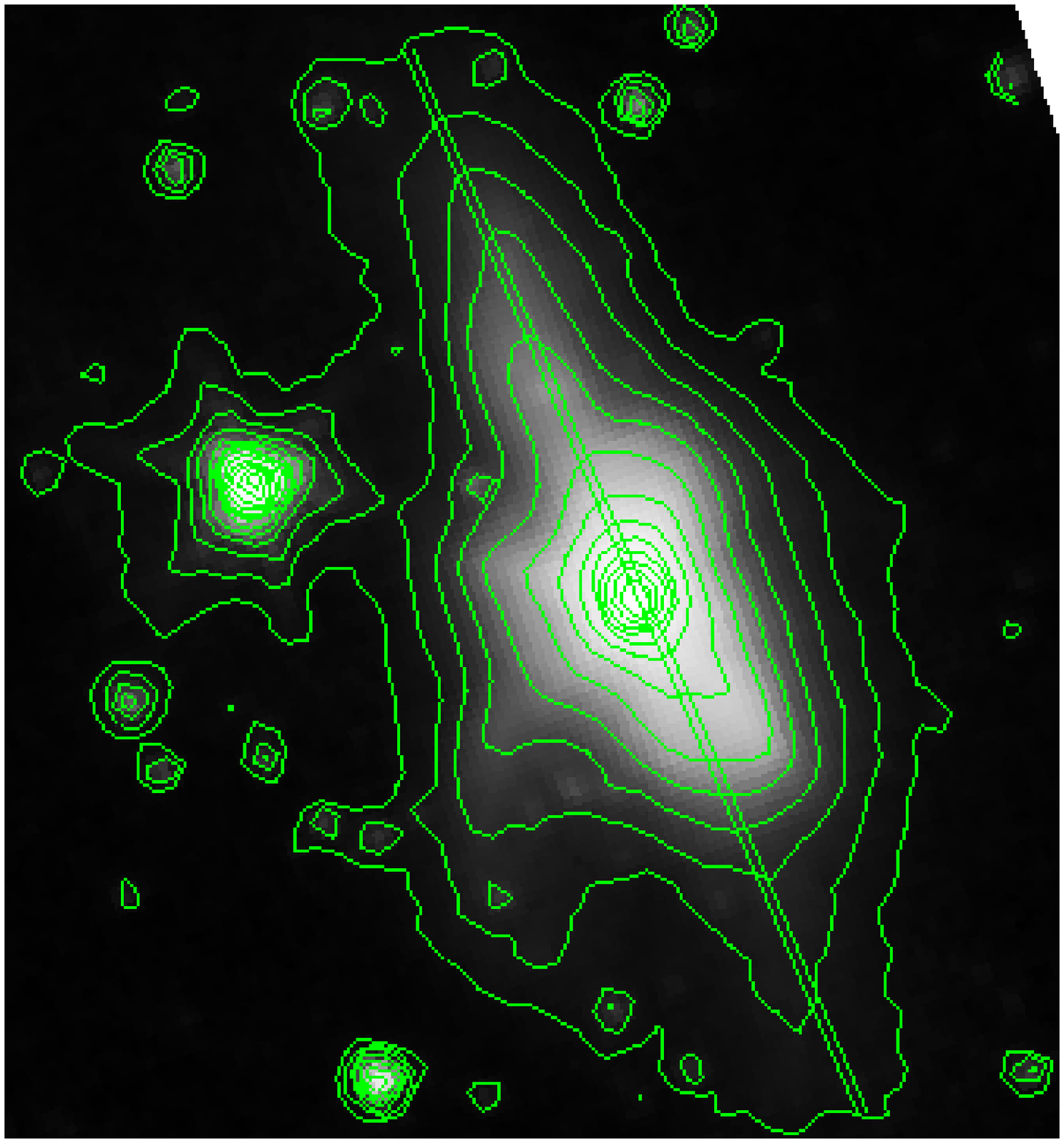}{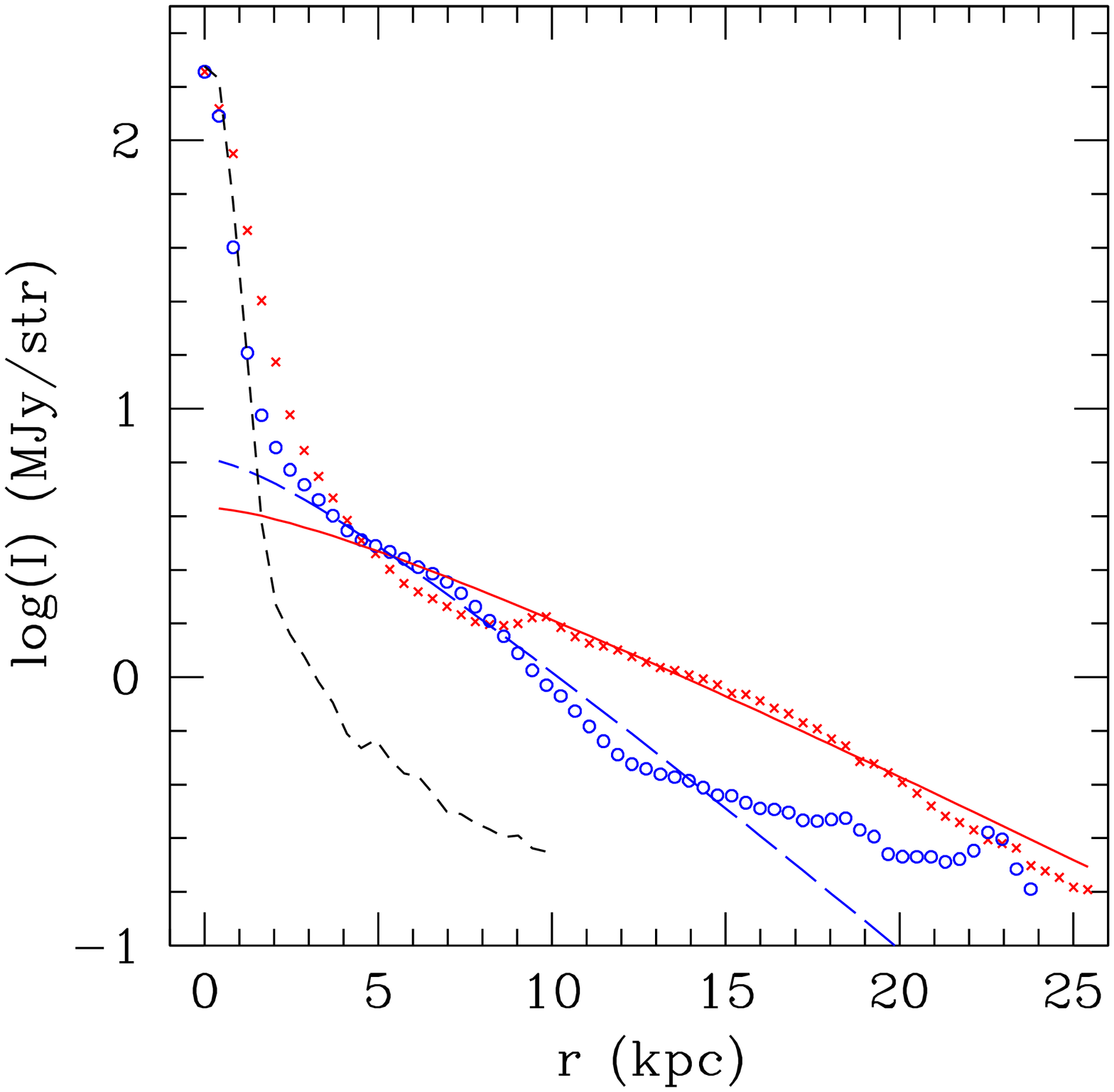}
\caption[]{Left: The two major axis cuts averaged to make the profile in Figure~\ref{fig:diskprofile} overlaid on the 3.6\,$\mu$m image with contours. Right: The surface brightness profile derived from the major axis cuts in the left panel. The red crosses are the northern side of the galaxy and the blue circles are the southern side of the galaxy. The outer parts look exponential and are fit with a theoretical surface brightness profile \citep{vanderKruit-Searle-1981}. These fits are the solid red line on the north side and the dashed blue line on the south side. These indicate the remains of a disk in this remnant. The black dotted line is the 3.6\,$\mu$m PRF. }
\label{fig:diskprofile}
\end{figure*}

It seems surprising that an active merger is well fit by an $r^{1/4}$ law, especially when its western half appears so disky. To check for signatures of a remnant disk, we averaged two radial cuts separated by one IRAC pixel along the major axis of the remnant. The locations of the cuts and the resulting profile are shown in Figure~\ref{fig:diskprofile}. On the northern side of the galaxy (crosses in Figure~\ref{fig:diskprofile}), the profile appears exponential beyond 10 kpc. We fit this portion of the profile with the edge-on galaxy surface brightness model of \citet{vanderKruit-Searle-1981} (solid line in Figure~\ref{fig:diskprofile}) which matches the data very well. The southern side is clearly not simply exponential. We attempt to fit the south profile beyond 5 kpc (the dashed line in Figure~\ref{fig:diskprofile}). This fit only agrees well for the portion of the profile from 5 to 15 kpc. This is not surprising given the truncated and asymmetric morphology of the southern side of the remnant. This, as all other aspects of this remnant, reflect a galaxy in transition, in the process of relaxing while some portions of the remnant disk remain. 

\section{Conclusions} \label{sec:conc} 

NGC 6240 is an object experiencing a rarely observed stage of evolution, showcasing complex combinations of processes. We add Spitzer-IRAC imaging and photometry to the active study of this object. We show that stars and dust have very different distributions in the remnant, both regionally and radially from the central AGN. We examine how outflows in the X-ray, H$\alpha$ and CO correlate with 8\,$\mu$m PAH emission. We show that though, when radially averaged, the stellar profile follows an   $r^{1/4}$ law surprisingly well, regions of the galaxy still show disk-like stellar profiles. All this supports the basic picture of a galaxy experiencing the effects of a major merger, which treats stars and gas/dust very differently, and can create a gas poor spheroid from a gas rich spiral galaxies. Though we can say little about the progenitors of this system, the sensitivity of IRAC data to low surface brightness emission adds detailed knowledge of the distribution of stars and gas in this remnant. This provides constraints for large samples of hydrodynamic simulations coupled with radiative transfer which, if they are able to match the characteristics of NGC 6240, may be able to determine its history and future. 

\section*{Acknowledgments}

 The authors gratefully thank Jiasheng Huang, Massimo Marengo and Joe Hora for assistance in reducing the IRAC data. We also thank Chris Mihos and Lars Hernquist for useful comments and discussions. We would like to thank the referee for helpful comments and guidance. This work is based on observations made with the Spitzer Space Telescope, which is operated by the Jet Propulsion Laboratory, California Institute of Technology under a contract with NASA. Support for this work was provided by NASA. MK is a member of the Chandra X-Ray Center,
which is operated by the Smithsonian Astrophysical Observatory under
NASA Contract NAS8-03060. Some of the data presented in this paper were obtained from the Multi-mission Archive at the Space Telescope Science Institute (MAST). STScI is operated by the Association of Universities for Research in Astronomy, Inc., under NASA contract NAS5-26555. Support for MAST for non-HST data is provided by the NASA Office of Space Science via grant NAG5-7584 and by other grants and contracts. This research has made use of the NASA/IPAC Extragalactic Database (NED) which is operated by the Jet Propulsion Laboratory, California Institute of Technology, under contract with the National Aeronautics and Space Administration.



\end{document}